\documentstyle[twocolumn,prl,aps]{revtex}
\begin{document}                                                
\draft
\title{A first order transition and parity violation 
in a color superconductor}
\author{Robert D. Pisarski$^{a}$ and Dirk H. Rischke$^{b}$}
\bigskip
\address{
a) Department of Physics, Brookhaven National Laboratory,
Upton, New York 11973-5000, USA\\
b) RIKEN BNL Research Center, Brookhaven National Laboratory,
Upton, New York 11973-5000, USA
}
\date{\today}
\maketitle
\begin{abstract} 
In cold, dense quark matter, 
quarks of different flavor can form Cooper pairs which are
anti-triplets under color and have total spin $J=0$. 
The transition to a phase where strange quarks 
condense with either up or down quarks is 
driven first order by the Coleman-Weinberg mechanism.
At densities sufficiently high to (effectively)
restore the axial $U(1)$ symmetry, then relative to the ordinary
vacuum, the condensation of up with down
quarks (effectively) breaks parity spontaneously.  
\end{abstract}
\begin{narrowtext}

By Cooper's theorem, a Fermi surface is destabilized by the presence of an
arbitrarily small attractive interaction \cite{super}.  
For quark-quark scattering, where quarks are triplets under
$SU(3)_c$ color, single gluon exchange
is attractive in the anti-triplet channel, and repulsive in the
sextet.  Thus it is reasonable to expect that at sufficiently
high densities, a Fermi sea of quarks forms a
quark-quark condensate in the color anti-triplet channel 
which spontaneously breaks the color gauge
symmetry.  In heavy ion collisions or quark stars,
``two plus one'' flavors of quarks contribute to the Fermi sea: 
up and down, which are very light, and strange, which is not as light.
In pioneering work, Bailin and Love \cite{bailin} showed that for
three colors and two flavors, 
color superconductivity breaks color, but not
flavor\cite{iwasaki,arw1,two,rg1}.  
Recently, Alford, Rajagopal, and Wilczek made the fascinating
suggestion that for three colors and 
three flavors, color superconductivity 
breaks both color and flavor\cite{arw2,rg2,abr,son}.

In this Letter we consider the nature of the color superconducting
phase in cold, dense quark matter, {\it assuming}
that condensation occurs in a channel with total spin $\vec{J}=
\vec{L}+\vec{S}=0$.  
Energetically, if condensation with $J=0$ can occur, it probably
is favored over $J\geq 1$.  For three colors and either two 
or three flavors, 
$J=0$ condensation is possible, and is indicated by 
most \cite{bailin,iwasaki,arw1,two,rg1,arw2,rg2,abr,son,dirk}
model calculations.

Under this assumption, we are able to make two general statements
about $QCD$ at nonzero quark density, where
the chemical potential $\mu \neq 0$, and zero temperature,
$T = 0$.  
First, that the transition to the color superconducting phase where 
strange quarks condense with either up or down quarks
is of first order.
Second, at very high densities,
where we can be certain that
the axial $U(1)$ symmetry is approximately restored, 
the condensate of up with down quarks spontaneously violates
parity.  Our arguments are general,
so we cannot say either how strongly first order this color superconducting
phase transition is, nor how dense quark matter must be in order to 
produce parity violation.  

A fermion-fermion condensate is naturally formed by pairing
at opposite edges of the Fermi sea, with momenta
$\vec{p}$ and $-\vec{p}$.  
In an ordinary superconductor, the charge of the 
electron-electron condensate is twice that of the electron.
Analogously, in a color superconductor
the quark-quark condensate has 
{\it two} color and {\it two} flavor indices\cite{multiple}:
\begin{equation}
\phi^{i j}_{a b}(\Gamma) \; = \;
q^i_a(-\vec{p})^T \, C \, \Gamma \; q^j_b(\vec{p}) \; ;
\label{e1}
\end{equation}
$q$ is the quark field, $q^T$ the Dirac transpose, 
$C$ the charge conjugation matrix, and $\Gamma$ a 
Dirac matrix.  The $SU(3)_c$ color indices
for a quark in the fundamental representation are $i,j=1,2,3$;
the flavor indices are $a,b$.  For now we ignore the color
and flavor structure, to consider how many spin zero
condensates there are.

An electron-electron
condensate with spin zero is formed by pairing spin up with
spin down.  Since electrons in an ordinary superconductor
are non-relativistic, 
spin decouples from the direction of motion,
and there is just one condensate. 

Massless quarks are eigenstates of chirality, and so come in two
types, left and right handed. 
As massless fields, their spin points either along
or opposite to the direction of
motion, corresponding to positive or negative helicity, respectively.
Thus massless quarks
have four types of spin zero condensates, two from chirality,
times two from helicity:
\begin{equation}
\Gamma = {\cal P}_{r,\ell} \times {\cal P}_\pm 
\; = \; \frac{1}{2} ( 1 \pm \gamma_5 ) \times
\frac{1}{2} ( 1 \pm \gamma_5 \gamma_0 \vec{\gamma} \cdot \hat{p} ) \; ;
\label{e2}
\end{equation}
${\cal P}_{r,\ell}$ is the projector for chirality, 
${\cal P}_\pm$ that for helicity.
It is an exercise in Dirac algebra to show that
these four condensates represent the 
pairing of right handed, positive helicity quarks
with themselves, {\it etc.} \cite{dirk}.

The total spin 
$\vec{J} = \vec{L} + \vec{S}$, so $J=0$ can arise
either from $L=S=0$ or $L=S=1$.  The signal for $L=S=1$ is
the appearance of $\hat{p}=\vec{p}/p$ in the helicity projector.

For a quark-quark condensate of massless quarks, Fermi statistics
requires that 
the condensate field is symmetric in the simultaneous interchange of
both color and flavor indices\cite{bailin,dirk}:
\begin{equation}
\phi^{i j}_{a b}(\Gamma) \; = \; + \, \phi^{j i}_{b a}(\Gamma) \; .
\label{e3}
\end{equation}
We use this constraint shortly to classify the possible
representations of color and flavor.  In an $SU(3)_c$ gauge
theory, $N$ flavors of 
massless quarks are invariant under the symmetries of local 
$SU(3)_c$ color and global $SU(N)_\ell \times SU(N)_r$ flavor.
Then $\phi^{i j}_{a b}({\cal P}_r {\cal P}_\pm)$ transforms under 
$SU(3)_c \times SU(N)_r$, and 
$\phi^{i j}_{a b}({\cal P}_\ell {\cal P}_\mp)$
under $SU(3)_c \times SU(N)_\ell$.

Parity flips both chirality and helicity, so these condensates
are not eigenstates of parity.  The condensates with
$\Gamma = \gamma_5$ and 
$\gamma_5 \gamma_0 \vec{\gamma}\cdot\hat{p}$ are parity
even, $J^P = 0^+$, while $\Gamma =1$ and 
$\gamma_0 \vec{\gamma}\cdot\hat{p}$ are parity odd, $J^P=0^-$.

At nonzero mass there are four new condensates \cite{bailin,dirk}.
In \cite{dirk} we show that two of these condensates are symmetric
under the interchange of color and flavor indices, as in (\ref{e3}),
while the other two are antisymmetric,
$\phi^{i j}_{a b}(\Gamma) = - \phi^{j i}_{b a}(\Gamma)$. 
The symmetric condensates are 
$\Gamma = \gamma_5 \gamma_0$ and $\vec{\gamma}\cdot\hat{p}$;
the antisymmetric condensates are 
$\Gamma =\gamma_5 \vec{\gamma}\cdot\hat{p}$ and $\gamma_0$.
In terms of parity, $\Gamma= \gamma_5 \gamma_0$ and 
$\gamma_5 \vec{\gamma}\cdot\hat{p}$ are $0^+$, while
$\Gamma =  \vec{\gamma}\cdot\hat{p}$
and $\gamma_0$ are $0^-$.  
A simple interpretation of these condensates suggests itself.
Relative to the quark momentum, there are three components of
spin: along the direction of motion, opposite, and transverse.
Times two for each parity gives a total of six condensates symmetric
in color and flavor.  Why massive quarks have two other condensates
which are antisymmetric in color and flavor is not apparent.

A quark mass mixes left and right handed fields, 
favoring the formation of a $0^+$ condensate\cite{arw1,two,rg1,arw2,rg2,abr}.
As instanton zero modes mix left and right handed fields,
the dynamical breaking of axial $U(1)$ symmetry also acts in the same
way.  We stress, however, that there are
{\it always} two kinds of helicity condensates \cite{toymodel}.  

We claim that to describe a phase transition to color superconductivity
in cold, dense quark matter, 
we only need to consider the condensate
fields, {\it ignoring} the gluons.
At one loop order, 
quarks screen gluons, generating a Debye
mass $m_D \sim g \mu$, where $g$ is the $SU(3)_c$
coupling constant \cite{hdl}.
We expect screening to occur at {\it any\/} nonzero density,
including the phase transition region.
If $m_D \neq 0$ at the
phase transition \cite{zerodebye}, as massive modes the gluons decouple over
large distances \cite{large}.

Consequently, in cold, dense quark matter
the phase transition to color superconductivity is governed by
an effective theory of the condensate fields in four dimensions \cite{hsu4d}.  
For scalar field theories in four dimensions,
the only possible infrared stable fixed point in the space of
coupling constants is the origin, so if of second order,
the transition is mean field with logarithmic corrections.
However, the transition could be driven first order by fluctuations through 
the Coleman-Weinberg mechanism \cite{coleman}.
Since the infrared stable fixed point is at the origin, this
possibility can be decided perturbatively, by computing
the $\beta$-functions for the scalar field theory to one loop order.  

The effective lagrangian is a sum of all terms which are invariant
under color and flavor rotations.  
There is always a nonzero density of instantons at the phase
transition; as shown in \cite{arw1},
they favor the condensation in 
the $0^+$ channel, and reduce the flavor
symmetry to $SU(N)_f$.
There are two $0^+$ condensate fields; in an
effective lagrangian they mix through mass terms, 
so that barring an accidental degeneracy, only one linear combination
of the two fields can become critical at a time.  

We are left to consider
the critical behavior of a single condensate field.  From 
(\ref{e3}), this field is symmetric in the interchange of
both color and flavor.  Therefore, 
the color and flavor representations must be either both symmetric,
or both antisymmetric.  From group theory, 
${\bf 2}\times {\bf 2} = {\bf 1}_a + {\bf 3}_s$ in $SU(2)$, while
${\bf 3}\times {\bf 3} = \overline{{\bf 3}}_a + {\bf 6}_s$ 
in $SU(3)$; ``$a$'' and ``$s$'' denote antisymmetric and symmetric
representations, respectively.

For the scattering of two quarks, single gluon exchange is attractive
in the color anti-triplet channel, $\overline{\bf 3}_a$, and repulsive in
the color sextet, ${\bf 6}_s$.  By asymptotic freedom, 
single gluon exchange is the dominant interaction at high density.
Thus we can assume that 
at all densities above some critical value, the color anti-triplet
channel is attractive.  (The
color sextet channel might become attractive for intermediate densities,
where the coupling is strong; if so, it is straightforward to
generalize our results.)

{\it Two flavors}: for $SU(3)_c \times SU(2)_f$ there are two allowed
representations, the color anti-triplet, flavor singlet
$(\overline{{\bf 3}}, {\bf 1})$ and the color sextet,
flavor triplet $({\bf 6}, {\bf 3})$.
The $(\overline{{\bf 3}}, {\bf 1})$ field is
a $U(3)$ vector, which is equivalent to 
an $O(6)$ vector.  The corresponding effective lagrangian has
one coupling constant, which is infrared free, and
so the transition can be of second order, with logarithmic corrections
to mean field exponents.

{\it Three flavors}: in $SU(3)_c \times SU(3)_f$ there are again
two possibilities, the color anti-triplet, flavor anti-triplet
$(\overline{{\bf 3}}, {\overline {\bf 3}})$ and 
the color sextet, flavor sextet $({\bf 6}, {\bf 6})$.
The color-flavor locking ansatz \cite{arw2,rg2,abr} contains both,
but only the $(\overline{{\bf 3}}, {\overline {\bf 3}})$ 
becomes critical \cite{mixing}.
The $(\overline{{\bf 3}}, {\overline {\bf 3}})$ 
field is a $U(3) \times U(3)$ vector.  
The corresponding effective lagrangian has two coupling constants;
the origin is infrared unstable \cite{unun}, implying that
the transition is driven first order 
through the Coleman-Weinberg mechanism \cite{det}.

{\it Two plus one flavors}: add a massive strange quark to massless
up and down quarks.  
The up and down condensates are as for two flavors.
Unlike massless quarks, massive strange quarks can condense
with themselves in the color anti-triplet channel with an
antisymmetric condensate \cite{dirk}, such as
$\Gamma = \gamma_5 \vec{\gamma}\cdot\hat{p}$.

Most interesting are the condensates of the strange quark with the
up and down quarks.  
There is no symmetry constraint on these
condensates, so
under $SU(3)_c \times SU(2)_f$, both the color anti-triplet, flavor doublet
$(\overline{{\bf 3}},{\bf 2})$ 
and the color sextet, flavor doublet $({\bf 6},{\bf 2})$ are allowed.
The $(\overline{{\bf 3}},{\bf 2})$ field is a 
$U(3)\times U(2)$ vector, with the transition driven
first order by the Coleman-Weinberg mechanism \cite{unun}.

In $QCD$, the important color anti-triplet condensates are the
$(\overline{{\bf 3}}, {\bf 1})$ field 
between up and down quarks, $\phi^i$, and the
$(\overline{{\bf 3}},{\bf 2})$ field  between strange and
either up or down quarks, $\widetilde{\phi}^i_a,\, a=u,d$.
Both of these fields condense at asymptotically
high densities.  Starting from low density,
due to the mismatch
between the Fermi surfaces of strange and nonstrange quarks,  
presumably first $\phi^i$, and {\it later}
$\widetilde{\phi}^i_a$, condenses; the former transition
may be of second order,
but the latter {\it must} be of first order.

So far we have ignored the chiral phase transition. 
Nonzero quark masses break chiral symmetry, so that
in $QCD$, if a chiral transition occurs, it cannot be
of second order, only
first (excluding a possible endpoint in the $\mu-T$ plane \cite{tri}).  
If there is a first order chiral transition for $\mu \neq 0$ and $T=0$,
it can either occur at the same point as where 
$\widetilde{\phi}^i_a$ condenses, or at a different point.
In either case, the transition where 
$\widetilde{\phi}^i_a$ condenses is of first order.

Note that in contrast to chiral symmetry,
color superconductivity is rather insensitive to 
nonzero, degenerate quark masses.   Single gluon exchange,
for example, remains attractive in the color anti-triplet channel;
while new condensates appear at nonzero mass, the condensates present
at zero mass should not change much.  It is important, however, that
different flavors have nearly degenerate masses, since only then are
the Fermi surfaces approximately equal.
Thus we suggest that in $QCD$, 
if there is a first order chiral transition at 
$\mu \neq 0$ and $T=0$, it 
might well occur at a different point from
where $\phi^i$ or $\widetilde{\phi}^i_a$ condense.
Model calculations with two massless flavors \cite{two}, 
however, do find
that $\phi^i$ condenses at the same point as a first order
chiral transition.

The patterns of color and flavor symmetry breaking can also be analyzed.
For two degenerate flavors,  a condensate for the 
$(\overline{{\bf 3}}, {\bf 1})$ field $\phi^i$ 
breaks $SU(3)_c$ to $SU(2)_c$ and leaves flavor unbroken \cite{bailin}.  
For three degenerate flavors, a condensate for the 
$(\overline{{\bf 3}}, {\overline {\bf 3}})$ field can break in two
distinct ways; one can show that color-flavor locking \cite{arw2},
where $SU(3)_c \times SU(3)_f$ breaks to $SU(3)$, 
{\it must} occur \cite{breaking}.  The case of two plus one flavors 
interpolates between these two limits.
If $\phi^i$ condenses and $\widetilde{\phi}^i_a$ not, the
breaking is that for two degenerate flavors, while once 
$\widetilde{\phi}^i_a$ condenses, at asymptotically high densities
$\phi^i$ and $\widetilde{\phi}^i_a$ 
combine to approach the case of three degenerate flavors.

Finally, we observe that
while there is always some breaking of axial $U(1)$, in practice
it becomes extremely small at high density.  
At high density, the breaking of axial $U(1)$ is reliably computable
by semi-classical means; for three massless flavors, at large $\mu$ the
density of instantons falls off like $\sim 1/\mu^9$.
Thus once 
semiclassical methods apply, we inescapably reach an ``instanton-free''
regime in which the axial $U(1)$ symmetry is effectively,
if not exactly, restored.  
This means that a $0^-$ condensate, such as
$q^T C q$, is as likely to form as the $0^+$ condensate
$q^T C \gamma_5 q$.  If axial $U(1)$ is an approximate symmetry, then
any mixture of these condensates is equally probable, 
so that condensation
spontaneously breaks not only axial $U(1)$, but also {\it parity}
(with respect to the vacuum at $\mu = T = 0$).
This is implicit in \cite{bailin}, noted by \cite{arw2}, and
stressed by \cite{rg1}.
This violation of parity may be large, since the axial $U(1)$
symmetry is nearly exact.  
The possibility of large parity violation at very high densities
may be important for both heavy ion collisions \cite{pol}
and quark stars \cite{stars}.

As the axial $U(1)$ is a continuous
symmetry, its breaking produces a Goldstone boson.
Including the quark masses turns this into a pseudo Goldstone
boson; unlike pions, however, 
its mass squared is not linear in the current quark
mass, but {\it quadratic}.  
To see this, note that 
even with a mass term, a discrete chiral symmetry still remains;
under $q^i_a \rightarrow i \gamma_5 q^i_a$, 
$\overline{q} q \rightarrow - \overline{q} q$ and
$\phi^i \rightarrow - \phi^i$.
If $m \rightarrow - m$,
then $m \overline{q} q$ and 
$m \phi^i$ are invariant under this discrete chiral symmetry;
but $m \phi^i$ is not invariant under color.
The simplest invariant term 
is $m^2 |\phi|^2$, which is quadratic in the quark mass.
The same result can also be shown directly from the gap equations.

Since the up and down quark masses are so much lighter than the strange,
the pseudo Goldstone boson for the spontaneous breaking
of axial $U(1)$ is almost entirely from the condensate of up with
down quarks; it is very light, with a mass 
of order $\sim 5 MeV$, and not $\sim 100 MeV$, as for
pions \cite{locking}.  

We thank S. Dawson, V. Emery, M. Gyulassy, D. Kharzeev,
V. N. Muthukumar, S. Ohta, K. Rajagopal, M. Stephanov, N. Samios,
E. Shuryak, and the referees for their help and insight.

\end{narrowtext}


\begin{references}
\bibitem{multiple}
At the outset one might ask, why a quark-quark 
condensate and not a color singlet six quark condensate with $J^P=0^+$?
A quark-quark condensate forms because of an infrared logarithmic
divergence in the scattering amplitude between two quarks near
the Fermi surface.  This divergence is cut off by a
quark-quark condensate, but not by higher quark condensates.
\bibitem{toymodel} 
This can be seen in a toy model \cite{dirk}, in which
colorless, flavorless, massless fermions $\psi$ interact with a scalar 
field $\chi$, of mass $M_s$, as $g_s \overline{\psi}\chi \psi$.  
Scalar exchange is attractive, so at nonzero fermion density
a fermion-fermion condensate forms.  Since the
interaction is not symmetric under axial $U(1)$, a $0^+$
condensate is favored; helicity then gives two types of
condensates, $\phi_\pm = 
\phi({\cal P}_r {\cal P}_\pm)-\phi({\cal P}_\ell {\cal P}_\mp)$.
As $M_s \rightarrow \infty$, there is only one condensate,
$\phi_+ = \phi_-$.
For any finite $M_s$, though, there are two distinct condensates,
$\phi_+ \neq \phi_-$.  Physically, this happens because 
as $M_s \rightarrow \infty$,
the interaction becomes point-like, so only processes with $L=S=0$
enter; at finite $M_s$, processes with $L=S=1$ can contribute, giving
$\phi_+ \neq \phi_-$. We note that previous studies
utilizing the Nambu
Jona-Lasinio model \cite{arw1}-$\!\!\!$\cite{rg2} only saw one
type of condensate, 
$\phi_+ = \phi_-$,
because the four-fermion interaction in that model
is point-like, corresponding 
to the limit $M_s \rightarrow \infty$.
In the toy model, the gap $\phi \sim \mu \; exp(-c_s/g_s^2)$.
In $QCD$, however,
we \cite{dirk}, and also Son \cite{son},
find instead that the gap depends upon the $SU(3)_c$ coupling constant
$g$ as $\phi \sim \mu \; exp(-c_{QCD}/\sqrt{g^2})$, where
$c_{QCD}$ is a constant.
\bibitem{zerodebye} 
If $m_D$ does vanish at the phase transition, gluons would contribute
to the critical
behavior. Then any color superconducting phase transition is driven
first order through the Coleman-Weinberg mechanism \cite{coleman}, as 
the asymptotically free gluonic couplings 
destabilize the infrared free scalar couplings.
\bibitem{mixing}
Denoting the 
$(\overline{{\bf 3}}, {\overline {\bf 3}})$ field by $\phi^i_a$,
and the $({\bf 6}, {\bf 6})$ field by $\chi^{i j}_{a b}$, $a,b=u,d,s$,
then because of
terms such as $det(\phi^\dagger) \phi^i_a \phi^j_b \chi^{i j}_{a b}$,
when $\phi^i_a$ develops a 
a vacuum expectation value, so will $\chi^{i j}_{a b}$.  
Because $\chi^{i j}_{a b}$ only condenses from its mixing with $\phi^i_a$,
though, the putative critical behavior is given solely by the $\phi^i_a$ field.
A similar phenomenon happens for $2+1$, but not for two, flavors.
\bibitem{det}
For three flavors the cubic term, $det(\phi)$,
is invariant under $SU(3)_c$ and $SU(3)_f$ rotations, but not under 
the $U(1)$ of
baryon number; thus it cannot appear in the effective lagrangian,
although the irrelevant operator $|det(\phi)|^2$ could.
Sch\"afer and Wilczek \cite{rg2} discuss using
$det(\phi)$, which has the quantum numbers of the $H$-dibaryon,
as a global order parameter.
For two plus one flavors, the analogous operator
is $\epsilon^{i j k} \epsilon^{a b} \phi^i \widetilde{\phi}^j_a
\widetilde{\phi}^k_b$ (see text for definitions).
\bibitem{locking}
Color-flavor locking \cite{arw2} also generates pseudo Goldstone masses 
which are linear in the quark mass.
However, these always
involve the strange quark mass, and so are much heavier, $\sim 100 MeV$.
\end{references}
\end{document}